\documentstyle[twoside,fleqn,espcrc2]{article}
\newcommand{\Z}{{\sf Z \!\!\! Z}}

\newcommand{\DPsi}{{\cal D}\Psi}
\newcommand{\DPsibar}{{\cal D}\overline{\Psi}}
\newcommand{\Psibar}{\overline{\Psi}}
\newcommand{\Deta}{{\cal D}\eta}
\newcommand{\Detabar}{{\cal D}\overline{\eta}}

\newcommand{\etabar}{\overline{\eta}}

\title{Fixed Point Actions for Lattice Fermions}

\author{W. Bietenholz\address{CBPF, rua Dr. Xavier Sigaud 150, 22290-180
        Rio de Janeiro RJ, Brazil} and
        U.-J. Wiese\address{H\"ochstleistungsrechenzentrum (HLRZ), c/o
        Forschungszentrum J\"ulich, 52425
        J\"ulich, Germany}\thanks{Based on two talks presented by the
        authors}}

\begin{document}

\begin{abstract}

The fixed point actions for Wilson and staggered lattice fermions are
determined by iterating renormalization group transformations. In both cases
a line of fixed points is found. Some points have very local fixed point
actions. They can be
used to construct perfect lattice actions for asymptotically free fermionic
theories like QCD or the Gross-Neveu model. The local fixed point actions for
Wilson fermions break chiral symmetry, while in the staggered case the
remnant $U(1)_e \otimes U(1)_o$ symmetry is preserved. In addition,
for Wilson fermions a nonlocal fixed point is found that corresponds to
free chiral fermions. The vicinity of this fixed point is studied in the
Gross-Neveu model using perturbation theory.

\end{abstract}

\maketitle

\section{Wilson Fermions}

The continuum limit of a lattice field theory is defined at a fixed point of
the renormalization group. The lattice models on a
renormalized trajectory emanating from the fixed point are free of cut-off
effects and hence have perfect lattice actions. Recently, Hasenfratz and
Niedermayer realized that perfect actions can be constructed explicitly for
asymptotically free theories \cite{Has93}. In addition, in the 2-d
nonlinear
$\sigma$-model the renormalization group transformation can be optimized such
that the fixed point action is extremely local. This is essential for numerical
simulations. The question arises if fixed point actions for
fermionic theories are local as well \cite{Wie93}. Since the fixed point of
an asymptotically free theory is close to the Gaussian fixed point this
question can be studied perturbatively, to lowest order even in the free
theory. The corresponding calculation for a free scalar field was done long
ago by Bell and Wilson \cite{Bel75}.

Let us consider free Wilson fermion fields $\Psibar$ and $\Psi$ with the
action $S[\Psibar,\Psi]$ on a hypercubic
lattice $\Lambda$, which is then blocked
to a lattice $\Lambda'$ of doubled lattice spacing. Then each point
$x' \in \Lambda'$ corresponds to a hypercubic block of $2^d$ points $x \in
\Lambda$ and each point $x$ belongs to exactly one block $x'$ (we denote this
by $x \in x'$).
 \setlength{\unitlength}{0.60mm}
 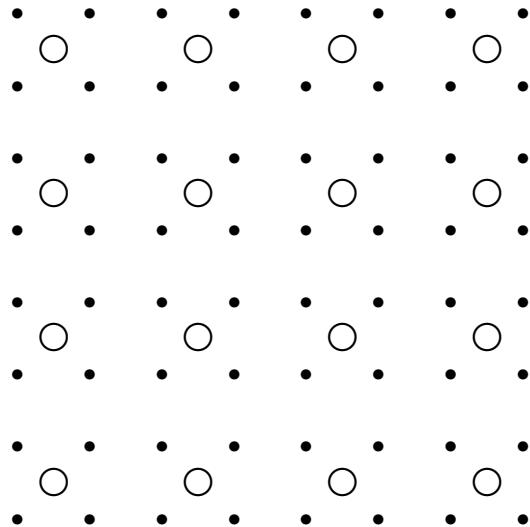
\begin{figure}[htb]
 \begin{minipage}{7.5cm}{}
 \begin{center}
 \begin{picture}(115,115)(10.0,10.0)
 \multiput(10,10)(16,0){8}{$\bullet$}
 \multiput(10,26)(16,0){8}{$\bullet$}
 \multiput(10,42)(16,0){8}{$\bullet$}
 \multiput(10,58)(16,0){8}{$\bullet$}
 \multiput(10,74)(16,0){8}{$\bullet$}
 \multiput(10,90)(16,0){8}{$\bullet$}
 \multiput(10,106)(16,0){8}{$\bullet$}
 \multiput(10,122)(16,0){8}{$\bullet$}
 \thicklines
 \multiput(19.5,19.7)(32,0){4}{\circle{6}}
 \multiput(19.5,51.7)(32,0){4}{\circle{6}}
 \multiput(19.5,83.7)(32,0){4}{\circle{6}}
 \multiput(19.5,115.7)(32,0){4}{\circle{6}}
 \end{picture}
 \end{center}
 \end{minipage}
 \caption{Blocking of a 2-d lattice.}
 \end{figure}
The block transformation is illustrated in fig.1.
On the blocked lattice we define new fermion fields
$\Psibar'$ and $\Psi'$ with an effective action $S'[\Psibar',\Psi']$, and we
perform a renormalization group step that leaves the partition function
unchanged
\begin{eqnarray}
\hspace{-7mm}
&&\exp(- S'[\Psibar',\Psi']) = \int \DPsibar \DPsi \exp(- S[\Psibar,\Psi])
\times \nonumber \\
\hspace{-7mm}
&&\exp\{- a \!\!\! \sum_{x' \in \Lambda'}
(\Psibar'_{x'} - \frac{b}{2^d} \!\! \sum_{x \in x'} \Psibar_x)
(\Psi'_{x'} - \frac{b}{2^d} \!\! \sum_{x \in x'} \Psi_x)\} \nonumber \\
\hspace{-7mm}
&&= \int \DPsibar \DPsi \Detabar \Deta \exp(- S[\Psibar,\Psi]) \times
\nonumber \\
\hspace{-7mm}
&&\exp\{\sum_{x' \in \Lambda'}
[(\Psibar'_{x'} - \frac{b}{2^d} \sum_{x \in x'} \Psibar_x) \eta_{x'} +
\nonumber \\
\hspace{-7mm}
&&\etabar_{x'}(\Psi'_{x'} - \frac{b}{2^d} \sum_{x \in x'} \Psi_x) +
\frac{1}{a} \etabar_{x'} \eta_{x'}]\}.
\label{step}
\end{eqnarray}
In the last step auxiliary fields $\etabar$, $\eta$ have been introduced.
For $a = \infty$ the renormalization group transformation corresponds
to a chirally invariant Grassmann $\delta$-function, whereas
for finite $a$ the
$\etabar_{x'} \eta_{x'}$ term breaks chiral symmetry explicitly. The
parameter $b$ is needed to renormalize the blocked fermion field.

Starting from a point on the critical surface and iterating the
renormalization group transformation one generates actions on coarser and
coarser lattices that may converge to a fixed point action $S^*[\Psibar,\Psi]$.
To find the fixed point we make the ansatz
\begin{eqnarray}
\hspace{-7mm}
&&S[\Psibar,\Psi] = i \sum_{x,y} \rho_\mu(x-y) \Psibar_x \gamma_\mu
\Psi_y + \nonumber \\
\hspace{-7mm}
&&\sum_{x,y} \lambda(x-y) \Psibar_x \Psi_y.
\label{ansatz}
\end{eqnarray}
We go to momentum space and integrate out $\Psibar$, $\Psi$ and $\etabar$,
$\eta$ to obtain the blocked action $S'[\Psibar',\Psi']$. At this point it is
convenient to introduce
\begin{equation}
\alpha_\mu(k)\!=\!\frac{\rho_\mu(k)}{\rho(k)^2 + \lambda(k)^2},
\beta(k)\!=\!\frac{\lambda(k)}{\rho(k)^2 + \lambda(k)^2}.
\end{equation}
After $n$ renormalization group steps one finds
\begin{figure}
\vspace{5.7cm}
\caption{The function $i \rho^*_1(z)$ for the fixed point action in coordinate
space at $a = 4$.}
\end{figure}
\begin{eqnarray}
\hspace{-7mm}
&&\alpha^{(n)}_\mu(k) = (b^2/2^d)^n \sum_l
\alpha_\mu(\frac{k + 2 \pi l}{2^n}) \times \nonumber \\
\hspace{-7mm}
&& \prod_\nu \left(\frac{\sin(k_\nu/2)}
{2^n \sin((k_\nu + 2 \pi l_\nu)/2^{n+1})} \right)^2, \nonumber \\
\hspace{-7mm}
&&\beta^{(n)}(k) = (b^2/2^d)^n \sum_l
\beta(\frac{k + 2 \pi l}{2^n}) \times  \\
\hspace{-7mm}
&& \prod_\nu \left(\frac{\sin(k_\nu/2)}
{2^n \sin((k_\nu + 2 \pi l_\nu)/2^{n+1})} \right)^2
+ \frac{1 - (b^2/2^d)^n}{a(1 - b^2/2^d)}. \nonumber
\label{fixed}
\end{eqnarray}
The summation is over integer component vectors $l$ with
$l_\mu \in \{1,2,3,...,2^n\}$. In the limit $n \rightarrow \infty$ one
approaches a nontrivial fixed point only if
\begin{equation}
(b^2/2^d)^n 2^n = 1 \, \Rightarrow \, b = 2^{(d-1)/2}.
\end{equation}
The exponent $(d-1)/2$
is the dimension of a free fermion field in $d$ dimensions, and the blocked
fermion field gets renormalized appropriately. At the fixed point one finds
\begin{eqnarray}
\hspace{-7mm}
&&\alpha^*_\mu(k) = \sum_{l \in \Z^d}
\frac{k_\mu + 2 \pi l_\mu}{(k + 2 \pi l)^2}
\prod_\nu \left(\frac{\sin(k_\nu/2)}{k_\nu/2 + \pi l_\nu} \right)^2,
\nonumber \\ \hspace{-7mm} &&\beta^*(k) = 2/a.
\label{fix}
\end{eqnarray}
Note that the result is independent of the parameter $r$ in the original
action, as it should since the Wilson term is irrelevant. One finds
a whole line of fixed points parametrized by $a$, and one may optimize $a$
such that the fixed point action is as local as possible. In $d=1$ the fixed
point action has only nearest neighbor couplings when $a=4$. This value
is close to optimal also in $d=2$.
In fig.2 the corresponding function $\rho^*_1(z)$ is displayed in coordinate
space.
Like in the scalar field case it is extremely local and very promising for
numerical simulations of perfect fermion actions.

\begin{figure}
\vspace{7.5cm}
\caption{The function $\rho^*_1(k)$ for the chirally invariant nonlocal fixed
point action.}
\end{figure}

For finite $a$ both the renormalization group transformations and the fixed
point actions break chiral symmetry and the fixed point actions are local.
For $a = \infty$, on the other hand, the renormalization group transformation
is chirally symmetric. As a consequence, there exists a chirally invariant
fixed point with $\rho^*_\mu(k) = \alpha^*_\mu(k)/\alpha^*(k)^2$,
$\lambda^*(k) = 0$, for which the Nielsen-Niniomiya theorem \cite{Nie81}
would suggest that the fermion spectrum is doubled. Fortunately,
this is not the case because the corresponding fixed point action is
nonlocal and the theorem does not apply. Hence, the resulting
continuum theory describes free chiral fermions. The function
$\rho^*_1(k)$ is shown in fig.3 for $d = 2$.
The poles at the boundary of the Brillouin zone give rise to the
action's nonlocality. SLAC fermions \cite{Dre76}
also have a nonlocal action which is known to cause problems in
perturbation theory \cite{Kar78}. For fixed point fermions, however, the
nonlocality should be acceptable, because it arises naturally due the
integration over the high momentum modes of the fermion field.

The real issue is to find a fixed point that describes
interacting chiral fermions. In particular, one should switch on a
small gauge coupling and investigate the vicinity of the nonlocal
fixed point of the free theory. Some time ago Rebbi \cite{Reb87} suggested
a lattice action for chiral fermions with a similar nonlocality, which
unfortunately
suffered from spurious ghost states \cite{Bod87,Cam87}. In fact, there is
a no-go theorem due to Pelissetto \cite{Pel88} that excludes lattice chiral
fermion constructions using certain nonlocal actions.
Still, we are optimistic that in case of the nonlocal fixed point action
the renormalization group is clever enough to circumvent these arguments,
although they certainly apply to generic man-made actions.

So far we have studied the vicinity of
the nonlocal fixed point in the Gross-Neveu model. The 4-Fermi interaction
is linearized by an auxiliary scalar field $\Phi$ with a Yukawa coupling
$y$. In momentum space the Yukawa coupling takes the form
\begin{equation}
\int d^2k_1 d^2k_2 \Psibar(k_1) \sigma(k_1,k_2) \Psi(k_2) \Phi(-k_1-k_2).
\end{equation}
The coupling $\sigma(k_1,k_2)$ is a matrix in Dirac space. Now
the renormalization group is iterated by also blocking the field $\Phi$ to
coarser and coarser lattices. At the fixed point of the coupled system one
finds to leading order in $y$
\begin{eqnarray}
\hspace{-7mm}
&&\sigma^*(k_1,k_2) = y \; \rho^*_\mu(k_1) \gamma_\mu
\sum_{l_1,l_2 \in \Z^2}
\frac{k_{1\nu} + 2 \pi l_{1\nu}}{(k_1 + 2 \pi l_1)^2} \gamma_\nu \times
\nonumber \\
\hspace{-7mm}
&&\frac{k_{2\rho} + 2 \pi l_{2\rho}}{(k_2 + 2 \pi l_2)^2} \gamma_\rho
\,\, \rho^*_\sigma(k_2) \gamma_\sigma \times \nonumber \\
\hspace{-7mm}
&&\prod_\lambda \frac{\sin((k_{1\lambda}+k_{2\lambda})/2)}
{(k_{1\lambda} + 2\pi l_{1\lambda} + k_{2\lambda} + 2\pi l_{2\lambda})/2}
\times \nonumber \\
\hspace{-7mm}
&&\frac{\sin(k_{1\lambda}/2)}{(k_{1\lambda} + 2\pi l_{1\lambda})/2}
\frac{\sin(k_{2\lambda}/2)}{(k_{2\lambda} + 2\pi l_{2\lambda})/2}.
\end{eqnarray}
Before attacking the gauge theory we will use this vertex function to check
if ghost states spoil the continuum limit in this nonlocal chirally invariant
formulation of the Gross-Neveu model. This investigation is in progress.

\section{Staggered Fermions}

An alternative formulation of lattice fermions uses staggered fermion fields.
This has the advantage that the cut-off theory has a remnant chiral symmetry.
The class of lattice fermions which is described by the
ansatz (\ref{ansatz})
does {\em not} include staggered fermions, since their
action is invariant only under translations by an even number of lattice
spacings. Staggered fermions can be treated analogously, but the
renormalization group transformations are different
and new properties of the fixed point actions arise.

As usual we view the staggered fermion variables $\bar \chi^i_x$,
$\chi^i_x$ as fields with pseudoflavors
$i = 1 + n_1 + 2 n_2 + ... + 2^{d-1} n_d$ ($n_\mu \in \{0,1\}$)
located at the $2^d$ corners of the hypercubes centered at points $x$
which form a lattice with spacing 2. Staggered fermions have various
symmetries, among them a $U(1)_e \otimes U(1)_o$ remainder of chiral
invariance, an analog of charge conjugation $C_0$, and shift
symmetries containing translations by one lattice spacing.
To be consistent with these symmetries
the renormalization group transformation has to be modified. At the end,
one wants to reconstruct Dirac spinors from the pseudoflavors.
Therefore it is important not to mix the corners of the hypercubes in the
renormalization group transformation. Kalkreuter, Mack and Speh have proposed
a suitable blocking scheme with block factor 3 (in general it must be odd)
\cite{Kal92}. One among $3^{d}$ block centers $x$ remains a block center after
the renormalization group step. We denote it by $x'$.
The $x'$ form a lattice of spacing 6. Each pseudoflavor builds its individual
block variable, and every $\chi^i_x$ on the
fine lattice contributes to exactly one $\chi'^i_{x'}$ on the coarse lattice.
The block transformation is illustrated in fig.4.
 \setlength{\unitlength}{0.60mm}
 \begin{figure}[htb]
 \begin{minipage}{7.5 cm}{}
 \begin{center}
 \begin{picture}(115,115)(10.0,10.0)
 \multiput(10,10)(16,0){8}{$\bullet$}
 \multiput(18,10)(16,0){7}{$\circ$}
 \multiput(10,18)(16,0){8}{$\diamond$}
 \multiput(18,18)(16,0){7}{$\ast$}
 \multiput(10,26)(16,0){8}{$\bullet$}
 \multiput(18,26)(16,0){7}{$\circ$}
 \multiput(10,34)(16,0){8}{$\diamond$}
 \multiput(18,34)(16,0){7}{$\ast$}
 \multiput(10,42)(16,0){8}{$\bullet$}
 \multiput(18,42)(16,0){7}{$\circ$}
 \multiput(10,50)(16,0){8}{$\diamond$}
 \multiput(18,50)(16,0){7}{$\ast$}
 \multiput(10,58)(16,0){8}{$\bullet$}
 \multiput(18,58)(16,0){7}{$\circ$}
 \multiput(10,66)(16,0){8}{$\diamond$}
 \multiput(18,66)(16,0){7}{$\ast$}
 \multiput(10,74)(16,0){8}{$\bullet$}
 \multiput(18,74)(16,0){7}{$\circ$}
 \multiput(10,82)(16,0){8}{$\diamond$}
 \multiput(18,82)(16,0){7}{$\ast$}
 \multiput(10,90)(16,0){8}{$\bullet$}
 \multiput(18,90)(16,0){7}{$\circ$}
 \multiput(10,98)(16,0){8}{$\diamond$}
 \multiput(18,98)(16,0){7}{$\ast$}
 \multiput(10,106)(16,0){8}{$\bullet$}
 \multiput(18,106)(16,0){7}{$\circ$}
 \multiput(10,114)(16,0){8}{$\diamond$}
 \multiput(18,114)(16,0){7}{$\ast$}
 \multiput(10,122)(16,0){8}{$\bullet$}
 \multiput(18,122)(16,0){7}{$\circ$}
 \thicklines
 \multiput(19.5,19.7)(24,0){5}{\circle{6}}
 \multiput(19.5,43.7)(24,0){5}{\circle{6}}
 \multiput(19.5,67.7)(24,0){5}{\circle{6}}
 \multiput(19.5,91.7)(24,0){5}{\circle{6}}
 \multiput(19.5,115.7)(24,0){5}{\circle{6}}
 \put(24.25,24.25){\framebox(38,38){\mbox}}
 \put(24.25,72.25){\framebox(38,38){\mbox}}
 \put(72.25,24.25){\framebox(38,38){\mbox}}
 \put(72.25,72.25){\framebox(38,38){\mbox}}
 \put(48.25,23.25){\dashbox{0.7}(38,40){\mbox}}
 \put(48.25,71.25){\dashbox{0.7}(38,40){\mbox}}
 \thinlines
 \put(46.75,48.25){\framebox(41,38){\mbox}}
 \end{picture}
 \end{center}
 \end{minipage}
 \caption{Blocking of a 2-d lattice consistent with the staggered
 symmetries.}
 \label{FigBlockLattice}
 \end{figure}
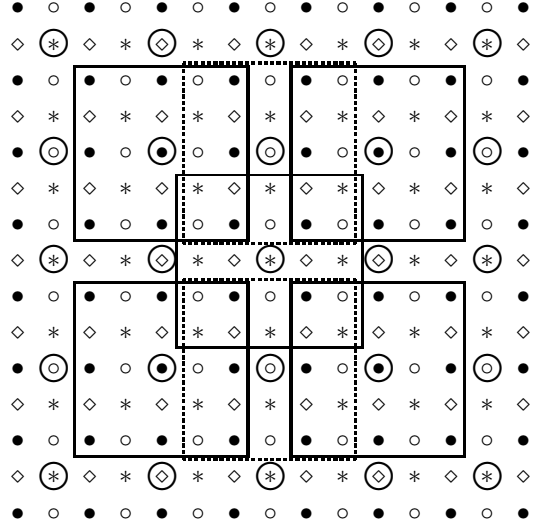
First we apply a $\delta$-function renormalization group transformation
using this blocking scheme
\begin{eqnarray} \label{deltargt}
\hspace{-7mm}
&&\exp(-S'[\bar \chi', \chi']) = \int {\cal D} \bar \chi {\cal D} \chi
\exp(-S[\bar \chi , \chi ]) \times \nonumber \\
\hspace{-7mm}
&&\prod_{x',i}
\delta (\bar \chi'^i_{x'}- \frac{b}{3^{d}} \sum_{x \in x'} \bar \chi^i_{x})
\delta ( \chi'^i_{x'} - \frac{b}{3^{d}} \sum_{x\in x'} \chi^i_{x} )
\nonumber \\
\hspace{-7mm}
&&= \int {\cal D} \bar \chi {\cal D} \chi
{\cal D} \bar \eta {\cal D} \eta \exp(-S[\bar \chi , \chi ]) \times
\nonumber \\ \hspace{-7mm}
&&\exp\{ \sum_{x' \in \Lambda'} \sum_i [(\bar \chi'^i_{x'} - \frac{b}{3^{d}}
\sum_{x \in x'} \bar \chi^i_{x}) \eta^i_{x'} + \nonumber \\
\hspace{-7mm}
&&\bar \eta^i_{x'}(\chi'^i_{x'} - \frac{b}{3^{d}} \sum_{x \in x'}
\chi^i_{x})]\}.
\end{eqnarray}
For each pseudoflavor we have defined coarse grained auxiliary Grassmann fields
$\bar\eta^i_{x'}$, $\eta^i_{x'}$.
Clearly this is analogous to the chirally invariant transformation
eq.(\ref{step}) for Wilson fermions with $a = \infty$.

Now we look for a general ansatz, analogous to (\ref{ansatz}),
which describes the action after a number of renormalization group steps
\begin{equation} \label{ansa}
S[\bar \chi,\chi] = i \sum_{x,y} \sum_{i,j} \bar \chi^{i}_{x}
\rho_{ij}(x-y) \chi^{j}_{y}.
\end{equation}
Here $\rho$ depends only on the distance $x-y$ between two hypercube centers,
because the action is invariant against translations by an even number of
lattice spacings.

The standard action as well as the renormalization group transformation
(\ref{deltargt})
respect various symmetries, which therefore also hold for (\ref{ansa}).
This specifies the form of the matrix $\rho$ as follows: the
$U(1)_e \otimes U(1)_o$ symmetry
implies $\rho_{ij} \neq 0$ only if the pseudoflavors $i$ and $j$ belong to
different sublattices (for one pseudoflavor the sum of lattice point
coordinates is even and for the other one it is odd), and $C_0$ invariance
requires $\rho_{ij}(x-y) = - \rho_{ji}(y-x)$. Shifts by one lattice spacing
yield further relations
among the remaining degrees of freedom; they relate sets of $2^{d-1}$
matrix elements.
As for
Wilson fermions it is natural to go to momentum space and to work with the
inverse matrix $\alpha(k) = \rho(k)^{-1}$.
In particular, for $d=2$ the various symmetries imply
\begin{eqnarray}
\hspace{-7mm}
&&\alpha(k) =
\left( \begin{array}{cccc} 0 & \alpha_1(k) & \alpha_2(k) & 0 \\
\alpha_1(k) & 0 & 0 & -\alpha_2(k) \\
\alpha_2(k) & 0 & 0 & \alpha_1(k) \\
0 & -\alpha_2(k) & \alpha_1(k) & 0 \end{array} \right). \nonumber \\
\hspace{-7mm} \,
\end{eqnarray}
In higher dimensions also couplings along certain space diagonals are
permitted by the symmetries; however, they are absent in the standard action
and it turns out that the renormalization group transformation does not
activate them either. Hence, in general one can parametrize the inverse
of $\rho(k)$ by $d$ functions $\alpha_\mu(k)$ with $\alpha_\mu(-k) =
- \alpha_\mu(k)$.

 After $n$ renormalization group steps we obtain
\begin{eqnarray}
\hspace{-7mm}
&&\alpha^{(n)}_\mu(k) = (b^2/3^d)^n \sum_l
\alpha_\mu(\frac{k + 2 \pi l}{3^n}) (-1)^{l_\mu} \times \nonumber \\
\hspace{-7mm}
&&\prod_{\nu}\left( \frac{\sin(k_\nu/2)}
{3^n \sin((k_\nu + 2 \pi l_\nu)/3^n 2)} \right)^2,
\end{eqnarray}
with $l_\mu \in \{1,2,3,...,3^n\}$.
The sign factor $(-1)^{l_\mu}$ arises because certain terms are
antiperiodic with respect to $k_\mu$. As for Wilson fermions
a nontrivial fixed point is reached only for one particular value of $b$; in
this case $b = 3^{(d-1)/2}$ in accordance with dimensional considerations.

It turns out that the $U(1)_e \otimes U(1)_o$ invariant fixed point action
is {\em local} as one infers from figs.5 and 6.
\begin{figure}
\vspace{7.5cm}
\caption{The function $i \rho^*_{12}(z)$ for the staggered fermion
fixed point in coordinate space ($d=2$).}
\end{figure}
\begin{figure}
\vspace{7.5cm}
\caption{The function $\rho^*_{12}(k)$ for the staggered fermion fixed point
action ($d=2$).}
\end{figure}
This is an important qualitative difference to the chirally invariant nonlocal
fixed point for Wilson fermions. In the staggered case there is no
contradiction to the Nielsen-Ninomiya theorem
since the $U(1)_e \otimes U(1)_o$ symmetry does not imply
full chiral invariance of the Dirac spinors that one can reconstruct.
Adding a mass term of the auxiliary fields (as it was done to optimize the
locality of the fixed point action for Wilson fermions)
would break the remnant chiral symmetry and would therefore destroy an
essential advantage of the staggered fermion formulation.
Still, we can improve the locality without loss of symmetry if we
add a kinetic term of the auxiliary
fields, again suppressed by a parameter $1/a$.
At the fixed point one obtains
\begin{eqnarray}
\hspace{-7mm}
&&\alpha^*_\mu(k) = 2 \sum_{l \in \Z^d}
\frac{k_\mu + 2 \pi l_\mu}{(k + 2 \pi l)^2} (-1)^{l_\mu} \times \nonumber \\
\hspace{-7mm}
&&\prod_{\nu} \left( \frac{\sin(k_\nu/2)}{k_\nu/2 + \pi l_\nu} \right)^2
+ \frac{9}{8a} \sin(k_\mu/2).
\end{eqnarray}
As long as no gauge fields are present one can still show that the partition
function remains invariant under renormalization group transformations.
In $d=1$ the fixed point action is an optimally local nearest neighbor
interaction when $a = 9/4$. Numerically it turns out that this
choice is also successful for $d=2$ as one sees in fig.7.
\begin{figure}
\vspace{7.5cm}
\caption{The function $i \rho^*_{12}(z)$ for the optimally local
fixed point ($d=2$, $a = 9/4$).}
\end{figure}
However, if we switch on a gauge
field this transformation is not allowed any more and we are restricted
to the $\delta$-function transformation ($a = \infty$) from above.
In summary we repeat that the fixed point actions we obtain for staggered
fermions are {\em local} and $U(1)_e \otimes U(1)_o$ {\em invariant}, and
are hence well suited for numerical simulations.

Next we intend to search for the fixed point action in the Gross-Neveu model
using staggered fermions.
Here the kinetic term of the auxiliary field can be used
and the location of the critical surface is obvious.
Since the Gross-Neveu model is asymptotically free, there is one (weakly)
relevant direction (to lowest order
it is marginal). When one determines this direction one obtains the
analog of the line of classical perfect actions in the bosonic
case \cite{Has93}.
Then one may adopt the concept of Hasenfratz and Niedermayer and
follow this (straight) line far enough to small correlation lengths
where numerical simulations become feasible.
If the simulated action is close to a perfect quantum action one can extract
continuum physics to a good approximation with only little numerical
effort. However, in the fermionic case it remains to be seen if an (almost)
perfect action can be simulated with the Hybrid-Monte-Carlo algorithm, or if
one is restricted to less efficient numerical methods.

We are indebted to P. Hasenfratz and F. Niedermayer for very interesting
discussions. We also like to thank T. Kalkreuter for providing us with a
copy of fig.4, and A. Pelissetto for explaining his no-go theorem.

\end{document}